\documentclass[12pt]{article}

\usepackage{times}

\usepackage[english]{babel}
\usepackage[T1]{fontenc}
\usepackage[latin9]{inputenc}
\usepackage{babel}
\usepackage{amsmath}
\usepackage{amssymb}
\usepackage{wasysym}
\usepackage{graphicx}
\usepackage{subfigure}
\usepackage{multirow}
\usepackage{color}

\usepackage{blkarray, bigstrut}
\usepackage{booktabs}

\newenvironment{sciabstract}{%
\begin{quote} }
{\end{quote}}

\title{Origin of the flat band in heavily Cs doped graphene}

\author
{N.~Ehlen$^{1\ast}$, M.~Hell$^{1\ast}$, G.~Marini$^{2}$, E.H.~Hasdeo$^{3}$,  R.~Saito$^{4}$, Y.~Falke$^{1}$,\\
  G.~Di~Santo$^{5}$, L.~Petaccia$^{5}$, G.~Profeta$^{2}$, A.~Gr\"{u}neis$^{1\ast}$\\
  \\
\normalsize{$^{1}$II. Physikalisches Institut, Universit\"at zu K\"oln,}\\
\normalsize{Z\"ulpicher Strasse 77, 50937 K\"oln, Germany}\\
\normalsize{$^{2}$Department of Physical and Chemical Sciences and SPIN-CNR,}\\
\normalsize{University of L'Aquila , Via Vetoio 10, I-67100 Coppito, Italy}\\
\normalsize{$^{3}$Research Center for Physics, Indonesian Institute of Sciences,}\\
\normalsize{Kawasan Puspiptek Serpong, Tangerang Selatan, 15314, Indonesia}\\
\normalsize{$^{4}$Department of Physics, Tohoku University, Sendai 980-8578, Japan}\\
\normalsize{$^{5}$Elettra Sincrotrone Trieste, Strada Statale 14 km 163.5, 34149 Trieste, Italy}\\
\\
\normalsize{$^{\ast}$To whom correspondence should be addressed;}\\
\normalsize{E-mail: ehlen@ph2.uni-koeln.de, hell@ph2.uni-koeln.de, grueneis@ph2.uni-koeln.de }\\
\normalsize{N.~Ehlen and M.~Hell have contributed equally.}
}

\begin{document} 


\baselineskip24pt


\maketitle


\begin{sciabstract}
  A flat energy dispersion of electrons at the Fermi level of a material leads to instabilities in the electronic system and can drive phase transitions. Here we introduce a method to induce a flat band in two-dimensional (2D) materials. We show that the flat band can be achieved by sandwiching the 2D material by two cesium (Cs) layers. We apply this method to monolayer graphene and investigate the flat band by a combination of angle-resolved photoemission spectroscopy experiment and the calculation. Our work highlights that charge transfer, zone folding of graphene bands and the covalent bonding between C and Cs atoms are at the origin of the flat energy band formation. The presented approach is an alternative route for obtaining flat band materials to twisting bilayer graphene which yields thermodynamically stable flat band materials in large areas.
 
\end{sciabstract}

\maketitle

\section*{Introduction}
An electronic energy dispersion with a narrow band width for a significant portion of the Brillouin zone (BZ) near the Fermi energy ($E_F$) is an intriguing electronic system~\cite{Kopnin2011,Leykam2018}. We refer to such a material, in accordance to literature~\cite{Kopnin2011,Leykam2018}, as a ``flat-band material'' even though the dispersion is usually not perfectly flat. Due to the singularity in the density of states (DOS) for the flat band, the material is unstable against opening of a gap near the $E_F$ that reduces the total energy. The instability can drive a phase transition of the system e.g. into a gapped superconductor, a charge-density wave, or magnetic ordering. Many theories predict that flat bands occur in the dice lattice~\cite{Sutherland1986}, the Kagome lattice~\cite{Syozi1951,Mielke1991,Bilitewski2018}, the Lieb lattice~\cite{Lieb1989} and the Tasaki lattice~\cite{Tasaki1992}. Some of the flat bands have recently been experimentally observed, e.g. the Lieb lattice~\cite{Drost2017,Slot2017}. Graphene related systems also provide ample opportunity for engineering flat bands. Theoretically, a carbon (C) Kagome lattice has been predicted~\cite{Zhong2016}, and recently, flat bands have been found in bilayer graphene~\cite{Marchenko2018} at $\sim$250~meV below the $E_F$ as well as in boron-doped graphene nanoribbons~\cite{Senkovskiy2018} at $\sim$1.5~eV below the $E_F$. However, since these flat bands are not located at the $E_F$, application of a gate volatage or chemical doping is needed to make the flat band relevant for the emergent ground state. It is thus important to induce the flat band near the $E_F$. Flat bands at the $E_F$ have been achieved in graphene systems by engineering the stacking order~\cite{faugeras16-rhombohedral,Pierucci2015,calandra17-rhombohedral,Henck2018}, the twist angle~\cite{Bistritzer2011,Cao2018} and the doping level~\cite{rotenberg10-extended}. Rhombohedrally-stacked trilayer graphene exhibits a flat band and, as a consequence, an antiferromagnetic ground state appears~\cite{Kopnin2011,faugeras16-rhombohedral,Pierucci2015,calandra17-rhombohedral,Henck2018}. Bilayer graphene where the two layers are twisted with respect to each other by a magic angle of $\sim$1.1$^\circ$ also exhibits a flat band close to the $E_F$~\cite{Bistritzer2011,Cao2018}. The discovery of both Mott insulating and superconducting phases in the $\sim$1.1$^\circ$ twisted bilayer graphene~\cite{Cao2018} highlights the physics of the flat band systems. One drawback of the rhombohedrally-stacked or the twisted bilayer systems is that they are fabricated from exfoliated flakes and hence can not be prepared deterministically and in large areas. Furthermore, the crystal structure of the twisted bilayer graphene is unstable against a rotation of the two layers, that brings the system back to the Bernal stacking order, which corresponds to the global energy minimum. Hence, the fabrication and characterization of large area systems that have flat bands at the $E_F$ is an important problem in condensed matter physics and materials science.

The present work introduces a new technique that is applied to induce a flat band at the $E_F$ of epitaxial graphene which is probed by angle-resolved photoemission spectroscopy (ARPES). The technique relies on the combined effects of heavy electron doping by excess cesium (Cs) / C stoichiometry, covalent C-Cs bonding and zone folding. Zone folding refers to the folding of graphene bands into a smaller BZ. Zone folding of energy bands occurs in the periodic potential of Cs atoms which is widely studied in the context of graphite intercalation compounds~\cite{Dresselhaus2002} and alkali-metal doped bilayer graphene~\cite{takahashi16-cs}. Recently, zone folding of graphene bands has been observed by oxygen intercalation under graphene/Ir(111)~\cite{osgood2019}. Zone folding leads to an increase in the number of bands in the folded BZ. For example, a $2\times 2$ zone folding increases the the number of bands by a factor of four compared with those in the unfolded band structure. Because of zone folding, the bands of doped graphene close to the $E_F$ occupy a central region of the BZ. Since the Cs band is also located around the BZ center for energies close to the $E_F$, hybridization between graphene and the Cs bands (i.e. a covalent coupling between C and Cs orbitals) is possible. It is noted that, without zone folding, there would be no hybridization close to the $E_F$ because the bands of graphene and Cs do not cross. According to the quantum-mechanical description, a coupling between C and Cs atoms causes anti-crossing of the energy bands. That is, the two dispersions do not actually cross, but disperse away from the hypothetical crossing point that would occur for no coupling. We will show in this work that hybridization and anti-crossing are key for obtaining a flat dispersion. A condition for the hybridization is that the alkali metal band is partially filled, i.e. the Cs atom should not be fully ionized. To ensure that the graphene layer is highly doped, we sandwich graphene in between two Cs layers. In fact, the doping levels we achieve are beyond the Lifshitz transition which occurs when the $E_F$ is higher than the energy of the conduction band at the $M$ point~\cite{rotenberg10-extended,Hell2018}.

Our work is closely linked to chiral superconductivity~\cite{Honerkamp2008,Nandkishore2012-chiral} and conventional superconductivity in graphene~\cite{Profeta2012,Chapman2016,ichinokura16-ca}. The former is a result of electron-electron interaction in the flat band~\cite{Honerkamp2008,Nandkishore2012-chiral}. The latter is driven by a large electron-phonon coupling (EPC) at the $E_F$ in alkali-metal doped graphene~\cite{Bianchi2010,Haberer2013,Fedorov2014,nikolay16-barium,takahashi16-cs,dima18-epc} in which a large electronic density of states (DOS) at the $E_F$ is favorable since the EPC constant can be written by $\lambda=N(0)D^2/M\omega^2$ where $N(0)$ is the DOS at the $E_F$, $D$ is the deformation potential and $M$ and $\omega$ are the mass and the phonon frequency, respectively~\cite{McMillan1968}. The critical temperature $T_c$ for conventional superconductivity depends on $\lambda$ and also on the renormalized Coulomb pseudopotential $\mu^*$ that likewise depends on $N(0)$~\cite{McMillan1968}. Thus, a decrease in the band width can lead to an increase of $T_c$. However, for the special case of superconductivity in alkali-metal doped graphene systems, not only a large $N(0)$ but also the EPC by additional coupling of phonons of graphene to alkali metal layer bands at the $E_F$ is required~\cite{Csanyi2005,Profeta2012,Smith2015}. Thus, alkali metal bands at the $E_F$ are necessary for superconductivity in graphene. This condition is equivalent to having a partially occupied alkali metal band (or interlayer state)~\cite{Csanyi2005,Profeta2012}. The two conditions (high doping level and partially occupied alkali metal band) apparently contradict each other - a high doping level (i.e. a flat band) would mean that all alkali atoms are fully ionized and thus the alkali derived band would be unoccupied whereas a low doping level would allow for the observation of the interlayer state but not the flat band. All previous works failed to engineer and characterize high-quality systems that allow for simultaneous observation of a flat band and the interlayer state. For example, Ref.~\cite{rotenberg10-extended} reports a flat band but no interlayer state and Refs.~\cite{Kanetani2012} and \cite{takahashi16-cs} find the interlayer state in bilayer graphene but their doping level is considerably lower than what is needed to occupy the flat band. In the present work we simultaneously observe a flat band at the $E_F$ and an alkali metal band by ARPES.

We also investigate the Raman spectrum of heavily doped graphene with a flat band. The doping strongly affects the Raman spectra of epitaxial graphene/Ir(111). The Raman $G$ band shifts down in frequency and assumes a Fano lineshape by alkali-metal doping~\cite{Hell2018}. The Fano lineshape of the $G$ band in alkali-metal doped graphene has been explained by quantum interference effects in the scattering pathways for vibrational and electronic Raman scattering (ERS)~\cite{Hell2018}. The relationship of $G$ band frequency and Fano asymmetry (expressed by the parameter $1/q$) versus carrier concentration has been probed for carrier concentrations up to the Lifshitz transition~\cite{Hell2018}. Here, a large and positive value of $1/q$ means a large Fano asymmetry of the observed Raman peak towards high wavenumbers. It was found that $1/q$ increases with carrier concentration up to $1/q=0.35$ for a carrier density at the Lifshitz transition~\cite{Hell2018}. The large $1/q$ is due to a resonance of electronic Raman scattering when the sample is doped slightly below the flat band~\cite{Hell2018}. However, although the Fano resonance and the flat band appear simultaneously in heavily doped graphene, the Raman response is not known and the relationship for the two phenomena is not clear.

\begin{figure}
	\centering
		\includegraphics[width=17cm]{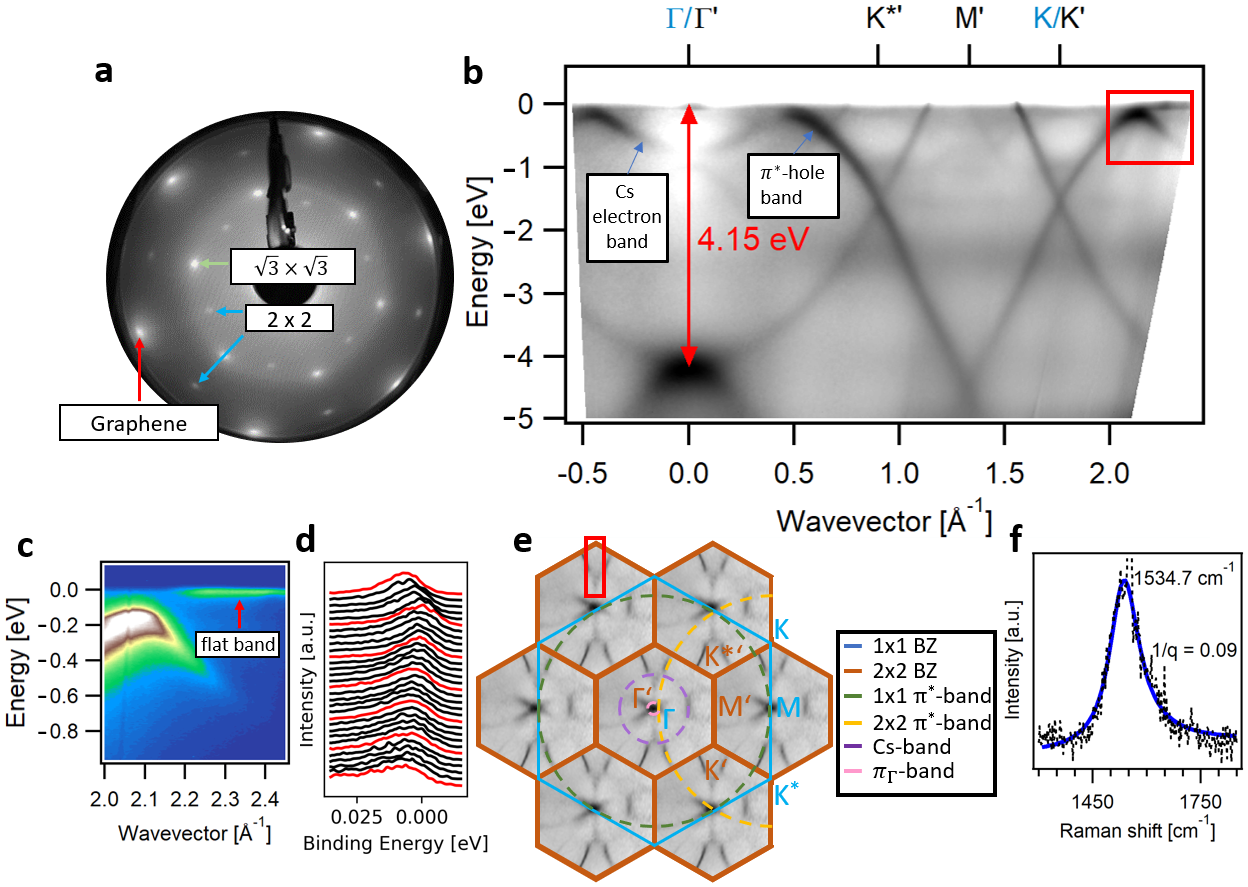}
	\caption{Characterization of Cs~2$\times$2/graphene/Cs~$\sqrt{3}\times\sqrt{3}$/Ir(111). (a) LEED pattern taken at an energy $E$=73~eV and a temperature $T=13$~K. (b) ARPES scan ($h\nu=31$~eV and $T=17$~K) along the $\Gamma KM$ directions. The region with the flat band is indicated by a red rectangle. The transition energy at the van Hove singularity of 4.15~eV is indicated. The high symmetry points of the original 1$\times$1 and the zone folded 2$\times$2 BZ are denoted in blue and black color, respectively. (c) Zoom-in to the region indicated by a red rectangle in (b) showing the flat band at the $E_F$. (d) Energy distribution curves (EDCs) for the flat band (stepsize 0.01{\AA}$^{-1}$) taken between 2.2~{\AA}$^{-1}$ (lower EDC) and 2.5~{\AA}$^{-1}$ (upper EDC). EDCs at every 0.05{\AA}$^{-1}$ are colored red. (e) Map at the $E_F$ with the $1\times 1$ and $2\times 2$ BZs. The map has been generated from a symmetrized azimuthal map taken in the first BZ. The $\pi^*$ bands of the original 1$\times$1 and the 2$\times$2 superstructure, the Cs band and the $\pi^*$ band from close to the $\Gamma$ point ($\pi_\Gamma$) are indicated. The red rectangle indicates a region along $\Gamma' K'$ containing the flat band. (f) Ultra-high vacuum Raman spectrum taken at $T=5$~K with a UV laser (325~nm) in a vacuum better than 1.0$\times 10^{-10}$~mbar.}
\end{figure}

\section*{Experimental results and discussion}
         {\bf Angle-resolved photoemission spectroscopy}\\
         The details of the preparation of monolayer graphene on an Ir(111) substrate are given in Refs.~\cite{Coraux2008,Hell2018}. A large amount of Cs corresponding to a thickness of approximately 30~{\AA} is evaporated onto graphene with a rate of about 1 {\AA} per 1~min (measured by a quartz micro balance) at room temperature. In these conditions, most Cs desorbs from the sample. The result of the 30~min long exposure to Cs vapor is that one Cs layer intercalates under graphene and the other Cs layer adsorbs on top of the graphene layer. We confirm that only negligible extra Cs atoms stick to the surface of the trilayer structure which is observed by sharp diffraction spots of graphene and Cs in the low energy electron diffraction (LEED) as shown in Figure 1a. The observed LEED pattern consists of three subpatterns: the hexagonal graphene lattice, a Cs $\sqrt{3}\times\sqrt{3}$-R30$^\circ$ pattern and a Cs $2\times 2$ pattern. From previous works that studied Cs interaction with graphene/Ir(111), it is known that the Cs~$\sqrt{3}\times\sqrt{3}$-R30$^\circ$ phase occurs for the Cs in between graphene and the metal substrate while the Cs~$2\times 2$ phase grows above the graphene~\cite{Petrovic2013,Hell2018}. Actually, when less Cs is deposited, we obtain the intercalated $\sqrt{3}\times\sqrt{3}$-R30$^\circ$ phase that we described above~\cite{Hell2018}.
         Thus, the graphene layer is encapsulated by one Cs~$\sqrt{3}\times\sqrt{3}$-R30$^\circ$ layer and one Cs~$2\times 2$ layer. Figure 1b depicts the energy band structure of this sample as is measured by ARPES along the high symmetry lines of the two-dimensional BZ. We observe zone folding of the electronic bands for the $2\times 2$ superstructure. It is important to note that the Cs~$\sqrt{3}\times\sqrt{3}$ order below the graphene does not induce zone folding. This can be explained by the screening of the Coulomb potential of Cs by the Ir which reduces the effective potential. On the other hand, for the Cs atoms on the top of graphene, the potential is sufficiently large to form the superstructure. The $2\times 2$ superlattice means that a folded BZ with half of the original reciprocal lattice vectors appears that corresponds to the 2$\times$2 unit cell. The high symmetry points of the original BZ which are relevant for the ARPES are $\Gamma$, $K$, $K^*$, and $M$ points. The relevant high symmetry points of the folded BZ are indicated by dashed symbols as $\Gamma '$,$K '$, $K^{*'}$ and $M'$ (see Figure 1b and 1e). Due to the zone folding, the $K^{*'}$ and $K'$ points of the zone folded BZ appear at the centers of $\Gamma K$ and $\Gamma K^*$ of the original BZ.
         
         Let us investigate the observed band structure to phenomenologically identify the origin of the flat band. From the ARPES, we see that there are two Dirac cones: one at $K$ which corresponds to a wavevector $k\sim 1.7$~{\AA}$^{-1}$ and the other one at $K^{*'}$ at $k\sim 0.8$~\AA$^{-1}$. The $\pi$ conduction bands are located above each of the two Dirac points in the ARPES spectra. In the present case, the conduction bands are partially occupied because of Cs doping. Each of the two $\pi$ conduction bands consists of two branches that disperse in opposite direction to each other with positive and negative curvatures which we will call $\pi^*$ electron and hole band, respectively. Around the $\Gamma$ point, we observe an electron band whose minimum is located at $0.7$~eV below the $E_F$. This electron band appears only by Cs doping. As we will show later by calculations, the electron band consists mainly of Cs 6$s$ states. We therefore call it Cs electron band. The Cs electron band hybridizes with the $\pi^*$ hole band whose maximum is at $\sim 0.5${\AA}$^{-1}$. The $\pi^*$ hole band and the Cs electron band are labelled in Figure 1b. Zone folding of the graphene bands is a key for their hybridization since the graphene bands cross the $E_F$ in the original $1\times 1$ BZ at wavevectors $\sim1.5${\AA}$^{-1}$ and  $\sim 2.2${\AA}$^{-1}$. Hybridization between the bands is equivalent to a covalent bonding of C and Cs atoms. The crossing of the $E_F$ by the Dirac cone bands of the $1\times 1$ BZ occurs far away from $\Gamma$, therefore a hybridization of graphene bands and Cs electron band close to the $E_F$ would not be possible. However, when we consider the zone folded Dirac cone, it can be seen immediately that the Cs electron band and the zone folded $\pi^*$ hole band can hybridize close to the $E_F$ near $\sim 0.5${\AA}$^{-1}$. The hybridization results in an anti-crossing of two branches. An extended flat band emerges as the higher energy branch of the anti-crossing located around $\Gamma$/$\Gamma'$ and $M$ points in the BZ. \\
         The fact that part of the conduction band is below the $E_F$, allows us to determine the transition energy at $\Gamma/\Gamma'$ to be $\approx$4.15~eV from ARPES (see Figure 1b).  Due to zone folding, the transition also appears at the $M$ point in the $1\times 1$ BZ. For optical measurements such as resonance Raman spectroscopy, the transition energy is expected to be reduced due to excitonic effects~\cite{Mak2011,Mak2014}. Thus we can resonantly excite heavily doped graphene and collect Raman spectra with good intensity if the laser energy is close to the value of 4.15~eV. In the following section we will show resonant Raman spectra that are measured using an ultraviolet laser with a wavelength of 325~nm (3.8~eV).
         Figure 1c depicts high resolution ARPES data in the region close to the $E_F$ where the $\pi^*$ hole band and the Cs electron band display the anti-crossing behaviour. The region appears at $k=2.2$~{\AA}$^{-1}$ as discussed before but due to zone folding it is equivalent to the region around $k=0.5$~{\AA}$^{-1}$. From Figure 1c, we can find evidence for a hybridization gap between the two branches near the $E_F$. The ARPES scan in Figure 1c shows the flat band in a range from $k=2.2$~{\AA}$^{-1}$ to a $k=2.4$~{\AA}$^{-1}$. However, considering the full 2D BZ, the flat band covers a much larger area as we will show later.
         Figure 1d shows the energy distribution curves of the flat band with a band width less than 10~meV. Notably, this band width is even smaller than the one observed for rhombohedrally stacked graphene where a band width of $\sim$ 25~meV has been observed~\cite{Henck2018}.
         
         In Figure 1e, we plot the 2D map of the ARPES intensity at the $E_F$ with highlighting the 2$\times$2 zone folding and the C and Cs derived energy bands. Due to zone folding, there are $\pi^*$ bands from the original $1\times 1$ BZ and the folded $2\times 2$ BZ that are indicated by green and yellow color, respectively. The $\pi^*$ derived Fermi surface contours and the Cs derived Fermi surface contours can be approximated well by circles (see the dashed circles in Figure 1e). ARPES matrix element effects are the reason why there are only segments with a strong ARPES intensity along the Fermi surface contours. Because of zone folding, the $\Gamma'K'$ and the $KM$ directions are equivalent. This allows us to observe the flat band segment shown in Figures 1b and 1c multiple times along the $\Gamma'K'$ directions of the zone folded BZs. One such segment is highlighted by a red rectangle in Figure 1e. Analysis of the observed Fermi surface requires consideration of charge carriers that occupy Cs and C derived bands. Using circles as approximations of the Fermi surface contours with C and Cs character, we obtain a total carrier concentration of $n=5.0\times 10^{14}$~cm$^{-2}$.

         Let us compare $n$ with the predicted carrier concentration for the 5/8 filling~\cite{Nandkishore2012-chiral}. In the nearest neighbor TB picture, the equi-energy contour at 5/8 filling consists of straight lines that connects adjacent $M$ points in the 2D BZ. For this case, we evaluate a large carrier concentration of $n=9.0\times 10^{14}$~cm$^{-2}$. However, due to the trigonal warping effect a much lower experimental value of $n=5.0\times 10^{14}$~cm$^{-2}$ is found~\cite{Ando1998,rsaito-book,Dresselhaus2002,CastroNeto2009}, i.e. the equi-energy contour in the 1$\times 1$ BZ does not connect adjacent $M$ points by straight lines but rather by trigonally warped curves. The trigonally warped equi-energy contour reduces $n$ compared without trigonal warping effect. Let us now break up the total charge carrier concentration into Cs and C states contributions. Evaluating the carrier concentration of the Cs band, we find that 0.54 electrons remain per 2$\times$2 unit cell (note: for a full charge transfer, we would have zero electrons remaining in the Cs band and thus the Cs band would appear above the $E_F$). The carrier concentration in the Cs band alone is $n=2.56\times 10^{14}$~cm$^{-2}$. 

{\bf Ultra-high vacuum Raman spectroscopy}\\
We have also performed Raman spectroscopy of an identically prepared Cs/graphene/Cs trilayer. Because of the high sensitivity of Cs doped graphene towards oxygen and moisture, these experiments were carried out in a homebuilt ultra-high vacuum (UHV) Raman system~\cite{Hell2018}. The sample quality was confirmed $in-situ$ by LEED where we have detected a diffraction pattern identical to Figure 1a.

The resonance Raman effect involves an optical transition from valence to conduction band~\cite{Saito2011-review,Ferrari2013-review}. The difference of the energy levels involved in this transition was evaluated from ARPES as 4.15~eV as discussed in the previous section. Since the transition energy is important for the observation of the present Raman spectrum of heavily doped graphene, we compare its value to the transition in pristine graphene. We have shown that the 4.15~eV transition occurs between valence and conduction bands at the $M$ point of the unfolded BZ. The transition energy at the $M$ point in the doped graphene sample is reduced when compared to pristine graphene~\cite{Hell2018}. Pristine graphene on Ir(111) has the $\pi$ valence band local minimum at the $M$ point with an energy of 2.8~eV (see Figure 3b of Ref.~\cite{Tusche2016}). The transition energy at the $M$ point can be expressed by the simplest tight-binding (TB) model in which we adopt one parameter, i.e. the transfer integral between the nearest C atoms, $t_{C-C}<0$ ~\cite{rsaito-book}. In the TB model, the valence and conduction bands are symmetric around the $E_F$ and the transition energy $E$ at the $M$ point is equal to $E=2|t_{C-C}|$~\cite{rsaito-book,CastroNeto2009}. When we use the experimentally observed valence band energy of 2.8~eV at the $M$ point~\cite{Tusche2016}, $2|t_{C-C}|=5.6$~eV which is 1.5~eV larger than the transition energy we observed by ARPES. A more accurate TB description~\cite{rsaito-book}, which also includes the effects of non-zero overlap matrix element $s$, yields the transition energy $E=2|t_{C-C}|/(1-s^2)$ at $M$. Using a typical literature value of $s\sim 0.1$ ~\cite{Bostwick2007,rsaito-book} we obtain a slight increase of $E$ in the order of one percent. For both cases of $s=0$ and $s\sim 0.1$, the calculated transition energy can not explain the ARPES result. Since the electron doping expands the C-C bonds~\cite{lazerri06-kohn}, $|t_{C-C}|$ is reduced. Further the energies of valence and conduction bands are affected by exchange and correlation energies that depend on the screening of the Coulomb interaction which can be provided by density functional theory. For a larger electron density, screening can be more efficient which contributes to the reduction of $t_{C-C}$. Ulstrup et al. compared ARPES spectra of doped graphene to density functional theory (DFT) calculations and found a reduction of the total $\pi$ band width upon doping~\cite{Ulstrup2016}. Thus the reduction of band width - and by extension the reduction of the transition energy - for Cs doped graphene can also be explained by the larger screening of the Coulomb interaction compared with that of pristine graphene.

Figure 1f depicts the Raman spectrum of the $G$ band for graphene in a Cs~2$\times$2/graphene/Cs~$\sqrt{3}\times\sqrt{3}$/Ir(111) structure. It can be seen that the $G$ peak is located at an energy of 1534.7~cm$^{-1}$. This energy is lower than was observed previously for doped graphene. The lower energy confirms larger doping of the present sample compared to previous works~\cite{brus2011-dopedgraphene,parret13-rubidium,Hell2018}. The peak shift can be understood well in terms of the existing theory describing the doping dependence of phonon modes in terms of lattice expansion and phonon self energy renormalization~\cite{lazerri06-kohn}. Using the relation between carrier concentration and $G$ band frequency that we have developed (figure 6a of Ref.~\cite{Hell2018}) to describe experiments of doped graphene on Ir(111), we correlate the peak position of the $G$ band phonon mode to a carrier concentration. For the experimentally observed $\omega_G=1534.7$~cm$^{-1}$ we would expect a carrier concentration of 4.6$\times$10$^{14}$~cm$^{-2}$. This is in good agreement to the ARPES derived carrier density of 5.0$\times$10$^{14}$~cm$^{-2}$. Interestingly, for the present system with larger carrier concentration, the Raman spectrum has a smaller Fano asymmetry of $1/q=0.09$ than for carrier concentrations below or at the Lifshitz transition that we previously measured~\cite{Hell2018}. As we show in the theory section, the dependence of $1/q$ on the carrier concentration reveals a sudden drop of the Fano asymmetry if the $E_F$ lies above the flat band. We explain this drop by the effects of electronic Raman scattering and Pauli blocking. The flat band phase is therefore characterized by Raman spectroscopy by a strongly renormalized $G$ band frequency of 1534.7~cm$^{-1}$ and by the value of the Fano asymmetry parameter $1/q=0.09$.

\begin{figure}
	\centering
		\includegraphics[width=15cm]{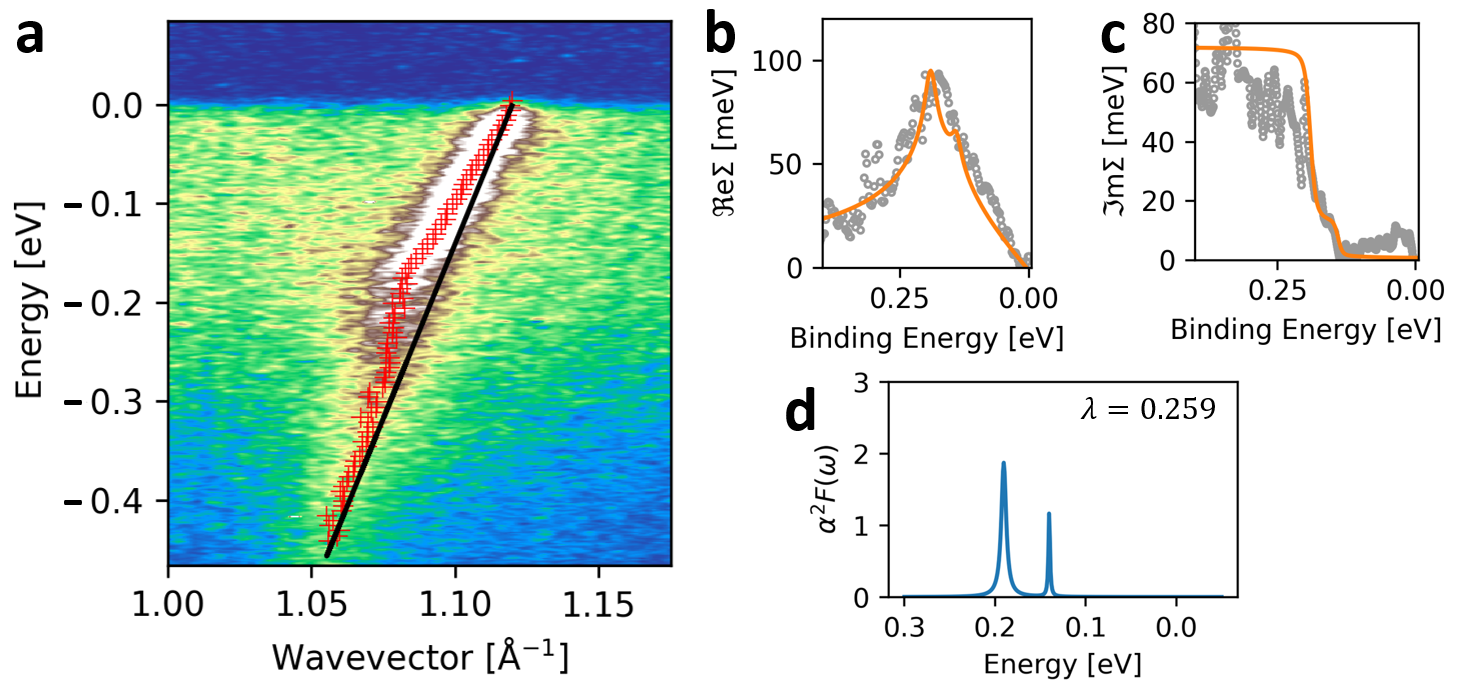}
	\caption{Analysis of the high resolution ARPES data in the region close to the Fermi level along $\Gamma K'^*$ direction. ARPES was taken at $h\nu=31$~eV and at $T=10$~K.  (a) ARPES scan in the vicinity of the Fermi wavevector and Fermi energy. The red crosses denote the ARPES maxima and the black line the bare band. (b) and (c) show the real and imaginary part of the self-energy, respectively. (d) the extracted Eliashberg function used for the calculation of the self-energy real and immaginary part calculations shown in (b) and (c) as solid line, respectively.}
\end{figure}
{\bf Analysis of electron phonon coupling}\\
Let us now turn to the analysis of EPC which results in a renormalized electronic dispersion that manifests as a kink in the measured spectral function close to the Fermi wave vectors along $\Gamma K$~\cite{Haberer2013,Fedorov2014,nikolay16-barium}. In Figure 2a, we show high resolution ARPES data taken in the vicinity of the EPC induced ``kink'' feature. The self-energy analysis of the kink is performed according to~previously established techniques~\cite{Haberer2013,Fedorov2014,nikolay16-barium}. Figures 2b and 2c depict the real and imaginary part of the self energy, respectively. The corresponding Eliashberg function extracted from the experimental data is shown in Figure 2d and has peaks at $\sim$200~meV and $\sim$150~meV for intravalley and intervalley EPC, respectively. Comparing the relative strengths of the two peaks with graphene doped to a lower carrier concentration, we find that the intravalley peak in the present case is dominant over the intervalley peak. We speculate that the relatively larger importance of intravalley EPC in the present case is related to the flat band which provides a large phase space for scattering with a phonon of a fixed energy. From integration over the complete Eliashberg function, we extract an EPC constant of $\lambda=2\int \frac{a^2F(\omega)}{\omega}d\omega=0.259$ along $\Gamma K$ which is the largest $\lambda$ in that crystallographic direction reported in doped graphene so far~\cite{Haberer2013,Fedorov2014,nikolay16-barium}. The Eliashberg function allows for discriminating the phonon origin of $\lambda$ by restricting the integration to a certain energy range. That is, if we integrate only over the high-energy peak that emerges due to the $\Gamma$ point phonon (identical to the Raman $G$ band) we obtain $\lambda_G=0.203$. Typically, the EPC in $KM$ directions is larger than the EPC along the $\Gamma K$ direction by a factor 2-3, see e.g. Refs.~\cite{Haberer2013,Fedorov2014,nikolay16-barium}. Thus we do expect also a record value of $\lambda$ along the $KM$ direction but unfortunately $\lambda$ cannot be reliably determined in this direction because the energy maximum of the branch along $KM$ is located at the phonon energy (see Figure 1c where the top of the band appears at 2.05~{\AA$^{-1}$} with 0.2~eV binding energy).  Note that the large $\lambda$ in the $\Gamma K$ direction is also nicely in-line with the large Raman downshift of the $G$ Raman mode as discussed in the previous section.

\section*{Origin of the flat band}
\begin{figure}
\centering
\includegraphics[width=15cm]{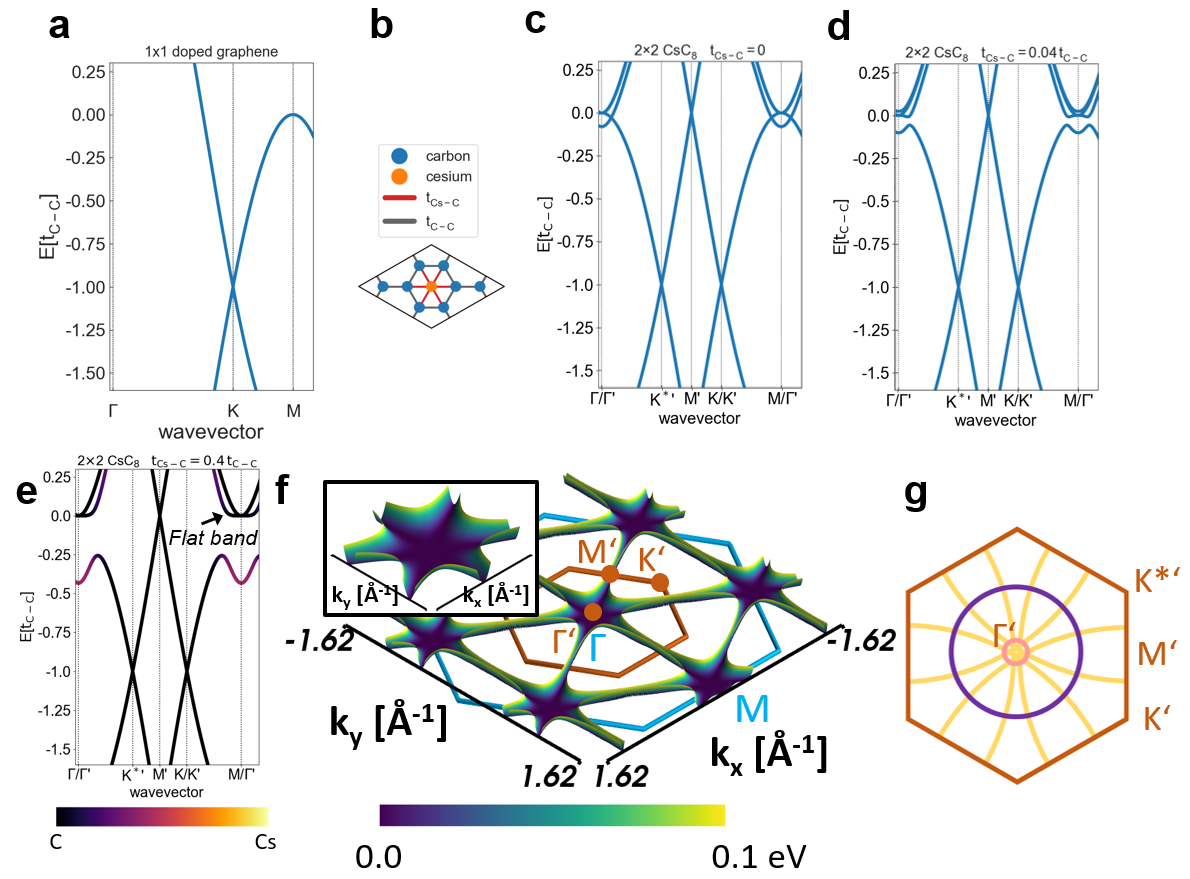}
\caption{(a) Tight-binding (TB) calculation of the band structure of electrostatically doped graphene in the original $1\times 1$ cell without Cs. (b) Geometry and TB hopping matrix elements for the $2\times 2$ Cs/graphene unit cell. The blue atoms are C and the orange atom is Cs. Red lines indicate Cs-C hopping ($t_{Cs-C}$) and black lines the C-C hopping ($t_{C-C}$) within the first unit cell. The six Cs-Cs hopping terms to the nearest neighbors are not indicated. (c) TB calculation of the electronic structure of $2\times 2$ Cs on graphene without Cs-C hybridization, i.e. $t_{Cs-C}=0$. (d,e) same as in (c) but with non-zero values of $t_{Cs-C}$. (e) The carbon (cesium) character of the band is indicated by black (purple) color. The flat band is marked by an arrow. (f) 3D plot of the band structure in a region between the $E_F$ and the $E_F+0.1$~eV using the parameters of (e). The brown (blue) hexagon denotes the $2\times 2$ ($1\times 1$) BZ. The inset shows a zoom of the region around the zone center. In all calculations all TB parameters except $t_{Cs-C}$ are kept constant. Parameters are expressed as multiples of $t_{C-C}$. We used $\epsilon_{Cs}=-t_{C-C}$, $\epsilon_{C}=t_{C-C}$ and $t_{Cs-Cs}=0.2t_{C-C}$ (see text for details on the TB model). All high symmetry points refer to the zone folded Brillouin zone shown in Figure 1e. (g) Experimentally derived Fermi surface contours in the zone folded $2\times 2$ BZ. The contours have been obtained by fits to the Fermi surface extracted from the ARPES experiments as shown in Figure 1e.}
\end{figure}

We study the effects of zone folding and hybridization between the Cs derived and graphene bands using a simple orthogonal nearest-neighbor TB model. The TB model considers only the upper Cs $2\times 2$ layer which causes the supercell formation since the bottom Cs $\sqrt{3}\times\sqrt{3}$ layer does not impose a sufficiently strong periodic potential as shown by ARPES. As we will show in the next section, the Cs atoms of the bottom layer are fully ionized and thus the bottom layer acts as a charge transfer layer. To accomodate the effects of charge transfer from the bottom layer into the TB calculation, we shift the on-site potential of graphene. Thus, the unit cell of our calculation includes one Cs~6$s$ and eight C~2$p_z$ orbitals in a $2\times 2$ unit cell with CsC$_8$ stoichiometry, defining a $9\times 9$ Hamiltonian matrix (see methods section). The parameters needed in our calculation are the two on-site energies for Cs and C orbitals (labeled $\epsilon_{Cs}$ and $\epsilon_{C}$) and the C-C, C-Cs and Cs-Cs hopping parameters indicated as $t_{C-C}$, $t_{C-Cs}$ and $t_{Cs-Cs}$, respectively. We start from a non-interacting monolayer graphene and switch on the $2\times 2$ potential and the Cs-C hybridization (i.e. the value of $t_{C-Cs}$) in steps. The band structure of doped graphene monolayer is shown in Figure 3a. The geometry of the unit cell of our TB calculation and the hopping parameters $t_{C-C}$ and $t_{C-Cs}$ are shown in Figure 3b. Let us now study the effect of zone folding and the effect of Cs-C hybridization on the electronic structure. Figure 3c depicts the bands calculated in a $2\times 2$ supercell but without Cs-C hybridization (indicated by a matrix element $t_{Cs-C}=0$). Due to zone folding, the Cs band and the graphene derived bands occupy the same region in the $E(k)$ plot. Since we set $t_{Cs-C}=0$ these bands do not interact with each other and they cross each other. The effect of non-zero $t_{Cs-C}$ becomes obvious by considering the regions in the BZ around $\Gamma$ and $M$ points. Figures 3d and 3e depict the calculated band structure of the $2\times 2$ system with values $t_{Cs-C}=0.04t_{C-C}$ and $t_{Cs-C}=0.4t_{C-C}$. It can be seen that a hybridization gap opens in the electronic spectrum in the crossing region between graphene and Cs derived bands. As a consequence of the anti-crossing, two branches emerge from that region. The higher energy branch forms an extended flat band at the $E_F$. Importantly, the extent of the flat region is a function of $t_{Cs-C}$ as can be seen by comparing Figures 3c and 3e. We observe that with increasing hybridization $t_{Cs-C}$, the flat band becomes more extended in the BZ. Figure 3e also shows the C and Cs character of the bands which we obtained from the eigenvector of the TB Hamiltonian (see methods). This calculation highlights that the flat band is derived from C states. Hence we conclude that it is inherited from the saddle point in the band structure at the $M$ point of the BZ of graphene. This conclusion is in agreement with the next section where first-principles calculations confirm the C character of the flat band. The simple TB model thus captures all the essential physics that were observed by ARPES. It highlights that the combined effects of zone folding and hybridization give rise to a flat band at the $E_F$. Importantly, it also means that the Cs derived bands must be occupied in order to form the hybridization. Partial occupation is the case if Cs is partially ionized. In the present experiment, the excess of Cs from the bottom layer ensures partial ionization of the upper Cs layer.

Let us now investigate the dispersion of the flat band in the 2D BZ. The 2D nature of the flat region is important for obtaining an instability in the electronic system because nesting of the electron wavevector is more efficient in an extended flat band. Figure 3f shows a 2D plot of the TB band structure close to the $E_F$. It can be seen that there are extended flat regions at $\Gamma$ and $M$ points of the $1\times 1$ BZ (or equivalently at the $\Gamma'$ point of the zone folded BZ). This theoretical result is in excellent agreement to the ARPES map shown in Figure 1e, where the extended flat regions around $\Gamma$ and $M$ points are visible dark regions. In the inset to Figure 3f, we show that the region around the zone center forms an extended flat band from which stripes of flat regions emerge in a star-like fashion with a sixfold symmetry. The direction of the stripes is along the $\Gamma M$ direction. To facilitate comparison of the TB calculated Fermi surface to the experimental Fermi surface, we plot the experimentally derived Fermi surface contours of the zone folded $2\times 2$ BZ in Figure 3g. The experimentally derived Fermi surface contours are obtained from fits of the ARPES intensity maxima by circular shapes as shown previously in Figure 1e. This procedure allows straightforward comparison of theory and experiment because it eliminates ARPES intensity variations due to matrix element effects. The Fermi surface that is shown in Figure 3g includes the zone folded Fermi surfaces of all neighboring BZs. Comparing the TB calculated contour with the contour derived from experiment along $\Gamma' M'$ we see observe good qualitative agreement. Theory and experiment deviate in the shape of the segment along $\Gamma' M'$. In the TB caculation (Figure 3f) it is straight but in the experiment (Figure 3g) it consists of two curved segements. We expect that this disagreement is due to trigonal warping which is not adequatly described in the nearest neighbor TB calculation. We expect that an improved TB calculation that includes more parameters will achieve quantitative agreement to the experiment.

Our approach might provide a solution to the intense efforts in materials engineering  to find new flat band materials and is not limited to small flakes but can produce large-area flat band materials. Such an approach is very much demanded in view of the current interest in flat bands in twisted bilayer graphene - a material that is not available in large areas. The simple nearest neighbor TB approach does not quantitatively describe the experiment, e.g. the experimentally observed band structure around the $M'$ point deviates from the TB calculation. More importantly, the flat part of the band structure in the experiment extends to a much larger part of the BZ, i.e. the simple nearest neighbor TB calculation underestimates the extent of the flat region. To that end, more parameters (hopping between neighbors that are further apart than one bond length) must be included.  A parameter-free description is given by first-principles calculations in the following section.

\section*{Theoretical results and comparison to experiment}
\begin{figure}
\centering
\includegraphics[width=17cm]{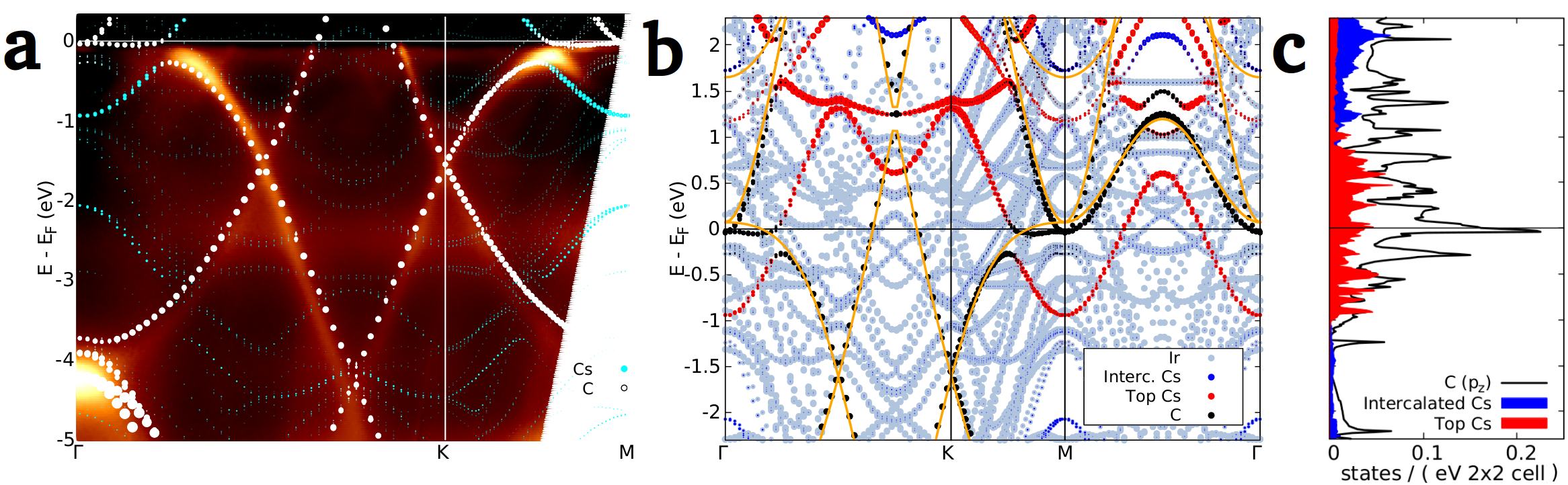}
\caption{(a) Density functional theory calculations (a shift of $-$130~meV was applied to match the experimental Fermi level) of 2$\times$2~Cs/graphene/2$\times$2~Cs along the $\Gamma KM$ directions (high symmetry points of the original 1$\times$1 BZ) overlaid on the experimental ARPES intensity. The color and the size of the dots indicate the atomic character of the corresponding projected eigenfunction of the Kohn-Sham eigenvalues (white for C and cyan for Cs-derived bands). (b) Calculated band structure projected on Ir (grey), C (black), intercalated Cs (blue) and Cs adatoms (red). The orange lines are calculations of electrostatically doped (i.e. without Cs) free-standing graphene in a $2\times 2$ supercell. (c) Corresponding partial density of states of C, intercalated Cs and top Cs.}
\end{figure}

{\bf Density functional theory calculations of the band structure}\\
First principles DFT calculations are performed by using a plane-wave pseudopotential approach~\cite{Kresse1996,Kresse1996a,Kresse1999} (see methods section for details). Since the $2\times 2$ and the $\sqrt{3}\times\sqrt{3}$ Cs superstructures are incommensurate to each other, the real structure is approximated by using $2\times 2$ Cs adsorbate layers on either side for simplicity. The two Cs layers are relatively shifted within in-plane direction by ${\bf a_1}+{\bf a_2}$, where ${\bf a_1}$ and ${\bf a_2}$ are the unit vectors of graphene. After Cs intercalation and subsequent adsorption we find that the graphene-Ir(111) perpendicular distance is 6.17~{\AA}, indicating a complete detaching of graphene from the substate. The  deposited Cs atoms are at distances of 2.97~{\AA} (for top Cs) and 3.01~{\AA} (for the bottom Cs) from the graphene layer. These distances are larger than what was obtained for Ba doped graphene in the same reconstruction~\cite{Tresca2018}. We will show later that these distances are important for the energetic position of the interlayer band. The calculated band structure of the Cs~2$\times$2/graphene/Cs~2$\times$2/Ir(111) system is plotted in Figure 4a, in the unfolded 1$\times$1 graphene BZ (note, the bands have been shifted down by 0.13~eV in order to match the experimental Fermi level). The comparison of the calculated results with the experimental ARPES spectrum reveals an impressive agreement between the theoretical and experimental bands. The effect of Cs on the band structure is crucial: apart from the obvious doping effect, the Cs~6$s$ orbital couples to the C~2~$p_z$ orbitals. Along the $KM$ direction, we observe a clear anti-crossing and the opening of a local hybridization gap of $\sim$200~meV. In the region of the gap opening, the orbital character of the band changes from C to Cs (see Figures 4a and c). This change in band character is in good agreeement to the TB-calculated change in band character as shown in the previous section. We also calculated the band structure of electrostatically doped graphene in the zone folded BZ (see Figure 4b - orange lines). The band structure of electrostatically doped graphene shows no such gap opening.

Going back to the Cs~2$\times$2/graphene/Cs~2$\times$2/Ir(111) system, we see that the anti-crossing results in an electron-like branch with Cs character centered at the $M$ point. The other branch of the band with avoided crossing disperses along the $E_F$ with a very narrow band width. We obtain a flat band that extends for one half of the $KM$ distance and quantitatively explains the origin of the flat band observed in the ARPES spectra. The flat dispersion gives rise to a van Hove singularity in the DOS; the corresponding peak is seen in Figure 4c. The Cs derived DOS is derived from the intercalated and adsorbed Cs layers. The intercalated Cs layer underneath graphene is almost completely ionized and acts as a charge transfer layer while the adsorbed Cs layer above graphene is only partially ionized and forms part of the Fermi surface. The DOS calculation corroborates full and partial ionization of the Cs layers below and above graphene, respectively. This can be seen in Figure 4c where the partial DOS of the upper Cs layer's 6$s$ orbital crosses the $E_F$ (red color area of the DOS of Figure 4c) whereas the partial DOS of the lower layer is localized mostly above the $E_F$. We note that the calculations have been performed for a $2\times 2$ intercalated Cs layer. We do not expect a qualitative difference for the experimentally observed $\sqrt{3}\times\sqrt{3}$ Cs structure apart from a slightly higher charge transfer for the $\sqrt{3}\times\sqrt{3}$ structure as a result of the higher Cs density. In the present calculation, the higher Cs density (and hence higher charge transfer) was accounted for by downshifting the band structure in energy by 0.13~eV.

\begin{figure}
  \centering
  \includegraphics[width=7cm]{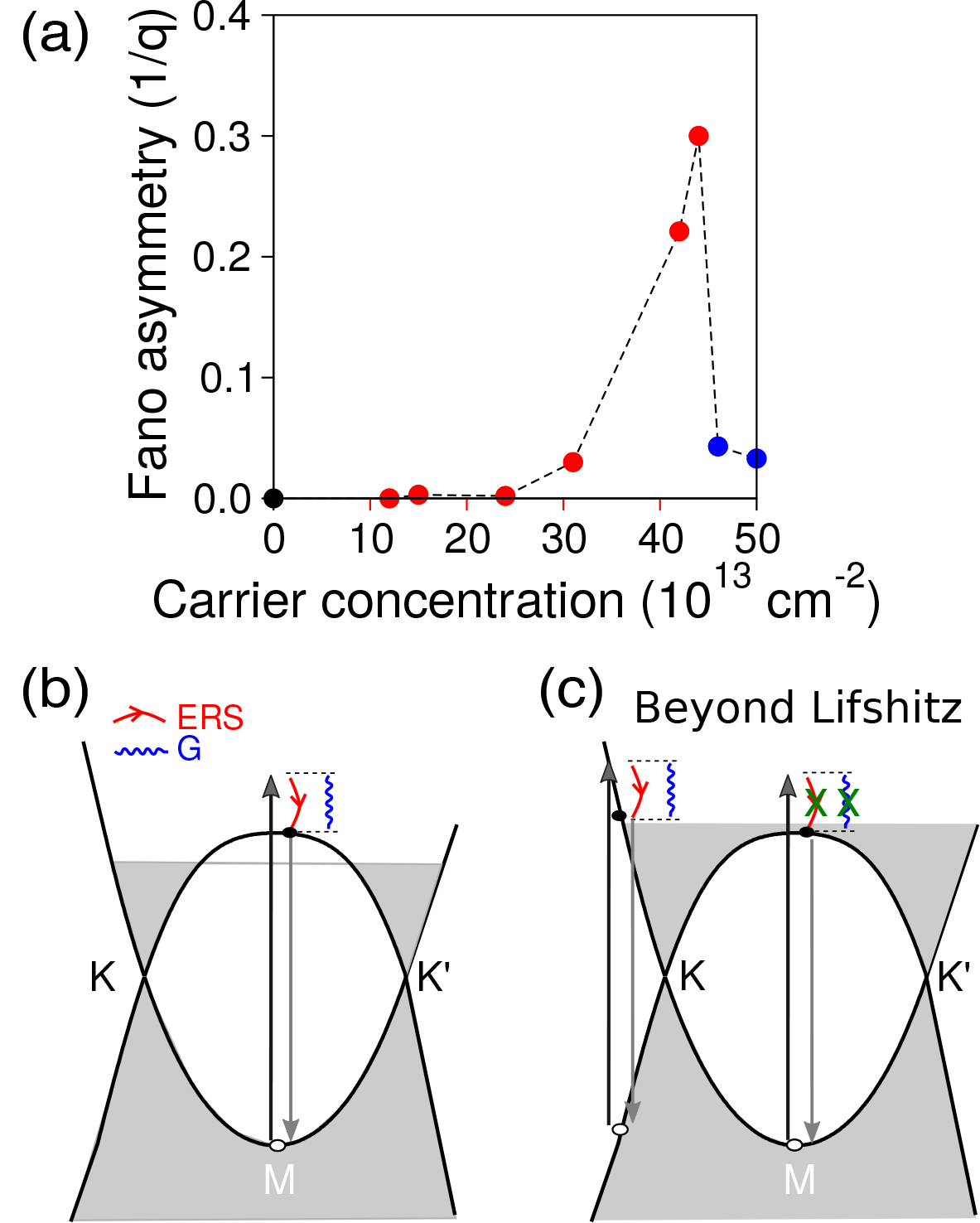}
	\caption{(a) Calculations of Fano asymmetry $1/q$ of the Raman $G$ band of doped graphene versus carrier concentration across the Lifshitz transition. The colors of the points carry the following meaning: black (pristine graphene), red (carrier concentration below the Lifshitz transition) and blue (carrier concentration above the Lifshitz transition). (b) and (c) sketches that depict the Raman interference between vibrational ($G$, blue) and electronic Raman scattering (ERS, red) before and beyond the Lifshitz transition. }
\end{figure}

{\bf Calculation of the Fano asymmetry of Raman spectra}\\
For calculations of the Raman spectrum of the $G$ band of doped graphene we have employed a previously developed model~\cite{Hasdeo2014-fano,Hell2018} which incorporates the effects of vibrational Raman scattering and ERS. Figure 5a depicts the calculated Fano asymmetry parameter $1/q$ versus charge carrier concentration, indicating an abrupt decrease to a value close to zero for carrier concentrations beyond the Lifshitz transition. The large Fano asymmetry in Raman for the $E_F$ just below the $M$ point conduction band maximum is a result of the interference of the $G$ band and ERS that are resonant with the van Hove singularity of the DOS (Figure 5b). Increasing the doping level hampers the resonant transition due to Pauli blocking (Figure 5c). As a result, the ERS intensity as well as the Fano parameter $1/q$ decrease. The observed sudden drop in $1/q$ can therefore be explained by the filling of the states at the $M$ point (or $\Gamma '$ point) as shown in the ARPES section. Thus the combined information of position and asymmetry can be used as a reference for Raman fingerprinting the electronic structure of a Cs/graphene/Cs trilayer. Note, that the calculated value of $1/q$ for the present carrier concentration is in excellent agreement to the observed Fano asymmetry. Theoretically, we obtain $1/q=0.05$ while the fit to the experimental $G$ band yields $1/q=0.09$ (see Figure 1f).
Interestingly, we did not observe evidence of zone folding of the phonon spectrum. For $\rm KC_8$, a new, low-intensity Raman active phonon mode that is derived from the $M$ point of the BZ has been reported~\cite{Smith2015,Dresselhaus2002}. We speculate that its weak intensity precludes its observation in epitaxial graphene since all Raman modes are weakened due to the metallic substrate.

\section*{Conclusions and outlook}
In the present work, we have used ARPES to directly proof the existence of a flat band in a Cs/graphene/Cs trilayer. TB calculations have revealed the mechanism of flat band formation. Two effects were found to be crucial: 1) zone folding of the graphene bands in a $2\times 2$ supercell and 2) hybridization of the zone folded graphene bands with the Cs metal 6$s$ bands at the $E_F$. Condition 2) also implies a partially ocuppied Cs band, i.e. an incomplete charge transfer of the upper Cs layer. The electronic structure obtained in the present system bears an interesting analogy to the electronic structure of the cuprate high temperature superconductors where the superconducting CuO$_2$ planes are next to charge reservoir layers. The role of the charge reservoir layer in the present case is played by the fully ionized Cs~$\sqrt{3}\times\sqrt{3}$ layer below graphene. The doped graphene and the upper Cs~$2\times 2$ layer act as charge transport layers. Thus the presented trilayer system and related structures may act as a test-bed for engineering superconductivity in 2D matter by chemical functionalization. The relevance to superconductivity is also clear from the large electron-phonon coupling that has been measured as ``kink'' feature in ARPES and as a phonon frequency renormalization in Raman spectroscopy. The presented system is relevant to both, chiral and conventional superconductivity because it hosts both, a flat band and a partially filled Cs band and has strong electron-phonon interaction.

Our TB calculations also reveal that the ``flatness'', i.e. the extension of the flat band in the 2D BZ can directly be controlled by the wavefunction overlap of the alkali metal $s$ orbital and the C~2p$_z$ orbital. The wavefunction overlap is given by the parameter $t_{Cs-C}$ in our TB calculations. This parameter is expected to change its value according to the type of alkali or earth alkali metal deposited and thus offers a wide tunability. We expect that Cs has a comparably large hybridization amongst the alkali metals because its outer electron occupies the 6$s$ orbital with large spatial extent. It therefore has a large overlap with the adjacent C~2$p_z$ wavefunctions. The other alkali atoms smaller.  We thus predict that the flat band is less extended in the 2D BZ for other $MC_8$ structures ($M$ being lithium, sodium, potassium or rubidium). From the variation of the parameter $t_{Cs-C}$ (Figure 2) it can be seen that the flatness increases for increasing $|t_{Cs-C}|$. In principle, the presented strategy to induce flat bands could be applied to any 2D material where the alkali metal order implies zone folding of the electronic structure of the host. It would be interesting to induce flat bands in the transition metal dichalcogenide (TMDC) family. TMDCs are known to also host ordered alkali metal intercalant and adsorbate phases. For example, Cs evaporated onto TiS$_2$ forms a Cs $2\times 2$ superstructure at certain Cs densities~\cite{Starnberg1997}. Another TMDC where the presented approach might work is the semiconductor MoS$_2$. The conduction band of MoS$_2$ has several flat segments at $Q$ and $K$ points and in the segment between $\Gamma$ and $M$ in the 2D BZ. These points give rise to maxima in the tunneling current in scanning tunneling experiments~\cite{Murray2019}. Moreover, akali metal evaporation onto the surface of MoS$_2$ results in a charge transfer to MoS$_2$ and the shift of the $E_F$ into the conduction band~\cite{Ehlen2018a}. A sufficiently large charge transfer and an ordered alkali metal adsorbate layer could cause hybridization of the MoS$_2$ conduction band with the alkali metal band and hence bring the flat segments of the conduction band down to the $E_F$.

\section*{Methods}
\subsection*{Angle-resolved photoemission spectroscopy and ultra-high vacuum Raman spectroscopy}
ARPES was performed at the BaDElPh beamline~\cite{Petaccia2009} of the Elettra synchrotron in Trieste (Italy) with linear $s$- and $p$- polarisation at $h\nu=31$~eV. The graphene/Ir(111) samples were prepared in-situ and measured in a vacuum better than $5\times 10^{-11}$mbar. Immediately after the synthesis, Cs deposition was carried out in one-shot in an ultra-high vacuum (UHV) chamber from commercial SAES getters with the sample at room temperature. The Fermi surface map from Figure 1 has been generated from a symmetrized azimuthal sweep of the first zone folded BZ taken at $T=13$~K. All ARPES measurements plotted  in Figure 1 are the sum of $s$ and $p$ polarization.

UHV Raman measurements were performed with the sample at $T=5$~K in the back-scattering geometry using a commercial Raman system (Renishaw) integrated in a homebuilt optical chamber~\cite{Hell2018a}, where the exciting and Raman scattered light were coupled into the vacuum using a HeCd laser with wavelength of 325~nm. The $20\times$ UV objective has a focal distance equal to 13~mm and an NA$=0.32$. The position of the laser on the sample was checked by a camera in the laser path. All spectra have been calibrated in position and intensity to the O$_2$ vibration at 1555~cm$^{-1}$ (see Ref.~\cite{Faris1997-oxygen}). The O$_2$ vibration is visible in the spectra due to the laser path outside the UHV. Sample preparation and Raman measurements were done in-situ and the sample was never exposed to air.

\subsection*{Tight-binding model of the flat band formation} 
To simulate the Cs $2\times2$/graphene structure we employed an orthogonal tight-binding Hamiltonian using nearest neighbor (nn) hopping for C-C, C-Cs and Cs-Cs bonds. The unit cell of the Hamiltonian consists of 8 C atoms (making up the $2\times 2$ supercell of graphene) and 1 Cs atom. The basis set is thus made up of 8 $p_z$-orbitals (one at each C site) and one $s$-orbital at the Cs site yielding a $9\times9$ Hamilton matrix
\begin{align}
\setlength\aboverulesep{0pt}\setlength\belowrulesep{0pt}
    \setlength\cmidrulewidth{0.5pt}
    H  = \begin{blockarray}{cccccccccc}
C1 & C2 & C3 & C4 & C5 & C6 & C7 & C8 & Cs & \\
\begin{block}{(cccccccc|c)c}
   &  &  &  &  &  &  &  &  & C1 \\
   &  &  &  &  &  &  &  &  & C2\\
   &  &  &  &  &  &  &  &  & C3 \\
   &  &  & H_{C-C} &  &  &  &  & H_{Cs-C}  & C4 \\
   &  &  &  &  &  &  &  &  & C5\\
   &  &  &  &  &  &  &  &  & C6\\
   &  &  &  &  &  &  &  &  & C7\\
   &  &  &  &  &  &  &  &  & C8\\
   \cmidrule(lr){1-9}
   &  &  & H_{Cs-C}^\dagger &  &  &  &  & H_{Cs- Cs}  & Cs\\
\end{block}
\end{blockarray}
\end{align}
that can be decomposed into an $8\times8$ Hamilton matrix $H_{C-C}$ describing the C-C intralayer hopping in the $2\times2$ unit cell, a $1\times8$ matrix $H_{C-Cs}$ describing the C-Cs interlayer interaction ($H_{C-Cs}^\dagger$ describes the hopping in the other direction) and a scalar function $H_{Cs-Cs}$ describing the Cs-Cs intralayer hopping. In the above Hamiltonian, C atoms are labelled as $C1$-$C8$ and the Cs atom as $Cs$.
The distance between nearest neighbors in the Cs layer is given by $d_{Cs-Cs}=2\sqrt{3}a_{C-C}$
with $a_{C-C}$ the C-C bonding distance. There are six nearest neighbor hopping directions in the lattice given by
\begin{align}
\begin{split}
    \Vec{\delta}^{Cs}_1 &= a_{C-C}(3,\sqrt{3})^T\\
    \Vec{\delta}^{Cs}_2 &= a_{C-C}(3,-\sqrt{3})^T\\
    \Vec{\delta}^{Cs}_3 &= a_{C-C}(-3,\sqrt{3})^T\\
    \Vec{\delta}^{Cs}_4 &= a_{C-C}(-3,-\sqrt{3})^T\\
    \Vec{\delta}^{Cs}_5 &= a_{C-C}(0,2\sqrt{3})^T\\
    \Vec{\delta}^{Cs}_6 &= a_{C-C}(0,-2\sqrt{3})^T.
\end{split}
\end{align}
The Cs-Cs hopping term can then be described by
\begin{align}
    H_{Cs-Cs} &= \epsilon_{Cs} + t_{Cs-Cs} \sum_{j=1}^6 \exp{(\mathrm{i}\Vec{k}\cdot\Vec{\delta}^{Cs}_j)} \\
    &= \epsilon_{Cs} + t_{Cs-Cs}\cdot h_{Cs-Cs}(\Vec{k})
\end{align}
with $t_{Cs-Cs}$ the hopping integral, $\epsilon_{Cs}$ the on-site potential of the Cs-lattice, and
\begin{align}
    h_{Cs-Cs}(\Vec{k})=(4\cos(3k_x a_{C-C})\cos{(\sqrt{3}k_ya_{C-C})} + 2\cos(2\sqrt{3}k_y a_{C-C})).
\end{align}
For the C-C and Cs-C hopping, it is helpful to define  the hopping directions
\begin{align}
\begin{split}
    \Vec{\delta}_1 &= a_{C-C}(1,0)^T\\
    \Vec{\delta}_2 &= a_{C-C}\left(-\frac{1}{2},\frac{\sqrt{3}}{2}\right)^T\\
    \Vec{\delta}_3 &= a_{C-C}\left(-\frac{1}{2},-\frac{\sqrt{3}}{2}\right)^T
\end{split}
\end{align}
and then define
\begin{align}
    f^\pm_j (\Vec{k}) = \exp(\pm\mathrm{i}\Vec{k}\cdot\Vec{\delta}_j)
\end{align}
which gives the $2\times2$ nearest neighbor Graphene Hamiltonian
\begin{align}
    H_{C-C} = \begin{pmatrix}
    \epsilon_C & 0 & tf^+_1 & 0 & tf^+_3 & 0 & tf^+_2 & 0\\
     & \epsilon_C & tf^+_3 & 0 & tf^+_1 & 0 & 0 & tf^+_2\\
      & & \epsilon_C & tf^-_2 & 0 & 0 & 0 & 0 \\
      & & & \epsilon_C & 0 & 0 & tf^+_1 &  tf^+_3 \\
      & & & & \epsilon_C& tf^-_2 & 0 & 0 \\
      & & & & & \epsilon_C & tf^+_3 & tf^+_1 \\
      & & & & & & \epsilon_C & 0 \\
      & & & & & & & \epsilon_C
    \end{pmatrix}
\end{align}
with $t=t_{C-C}$ the hopping integral and $\epsilon_C$ the on-site potential of the C lattice. The lower triangle can be constructed from the upper triangle by using the Hermitian condition of Hamiltonians.
The C-Cs interlayer hopping can be described by
\begin{align}
    H_{C-Cs} = \begin{pmatrix}
        0 \\
        t_{C-Cs} f_2^- \\
        t_{C-Cs} f_1^+ \\
        t_{C-Cs} f_3^- \\
        t_{C-Cs} f_3^+ \\
        t_{C-Cs} f_1^- \\
        t_{C-Cs} f_2^+ \\
        0
    \end{pmatrix}
\end{align}
with $t_{C-Cs}$ the hopping integral for nearest neighbor C-Cs hopping. The band structures in Figure 2 have been calculated by solving $H({\bf k}){\bf c}({\bf k})={\bf c}({\bf k})E({\bf k})$ where ${\bf c}({\bf k})$ and $E({\bf k})$ are the eigenvector and eigenvalue, respectively. The Cs or C character of the bands (Figure 2e) was calculated from the value of the corresponding eigenvector components.

\subsection*{Computational details of the DFT calculation and modelling of the system} 
First-principles density functional theory calculations were performed to describe the Cs doped graphene system on an Ir(111) surface. We used the pseudopotentials approximation for the electron-ion interaction and a plane-wave expansion of the Kohn-Sham wavefuncti1ons as implemented in the \textit{VASP} package~\cite{Kresse1996,Kresse1996a,Kresse1999}. Generalized gradient approximation (GGA) in the Perdew, Burke and Ernzerhof (PBE) formulation has been adopted for the exchange-correlation potential. A 400~eV cutoff for the plane waves basis set and $14^2$ $\Gamma$ centered \textbf{k}-point grid with a gaussian smearing of 0.1 eV have been employed. Long range van-der-Waals interaction (important to describe graphene-substrate interactions)  have been included by the Grimme's semiempirical correction (DFT-D2) to the functional~\cite{Grimme2006}. Spin-orbit coupling has been self-consistently taken into account.

We modelled the system using a 2$\times$2 graphene unit cell on-top of an Ir(111) terminated surface. Due to the lattice mismatch between graphene and the ideal Ir(111) surface we fixed the in-plane lattice parameter at the graphene equilibrium value (2.47 {\AA}), thus straining the Iridium in-plane lattice constant by $\approx$ 9 \%. The out-of-plane distance between two Iridium layers has been fixed to its bulk value, using the Ir bulk lattice constant (3.84 {\AA}). Four Ir layers and 25 {\AA} of vacuum have been used in the calculation adding dipole correction to account for the inequivalent top and bottom surfaces of the slab. We treated Cs doping adding one Cs atom per unit cell below the graphene layer as intercalant to detach graphene from the substrate and additional Cs atoms on-top of the graphene layer, both in the hollow sites of the C hexagons. With this unit cell, we found that the lowest energy configuration for the adsorbed Cs atoms (above and below graphene) is the one with  Cs atoms occupying the center of the two inequivalent  hexagons of the 2$\times$2 unit cell. The positions of the C and Cs atoms have been relaxed until the forces on the atoms are less than 0.01 eV/{\AA}, while the Ir atoms were fixed to their bulk sites.

Interestingly, both the position and the dispersion of the flat band are slightly affected by spin-orbit coupling induced by the Ir substrate. Although spin-orbit coupling in graphene is negligible, the presence of the Ir substrate induces a relevant reconstruction of the band structure around the $\Gamma$-point. The calculated  band structure without the inclusion of the spin-orbit interaction shows that the Ir-derived hole-pocket at the $\Gamma$-point overlaps with graphene $\pi$ bands and is strongly hybridized with them. In particular, we reveal the presence of a flat band at the $E_F$ at the $M$-point.
When we switch on the spin-orbit interaction in the calculation, we get a significant change for the Ir states. In particular, at the $M$ point, spin-orbit coupling induces a gap-opening: the Ir bands become lower in energy by strongly reducing the hybridization with graphene bands. Thus the Ir band becomes fully occupied with a band maximum at 175~meV below the $E_F$. The Ir band shows a Rashba type splitting of 0.079~{\AA}$^{-1}$ which is in agreement with that reported in Ref.~\cite{Varykhalov2012}. This value can be estimated by the energy position of the surface state of Ir(111)\cite{DalCorso2015}.

\subsection*{Calculation of the Raman spectrum}
We have performed calculations of the Raman spectrum according to previously developed model~\cite{Hasdeo2014-fano,Hell2018}. Raman intensity as a function of Raman shift $\omega_{\rm s}$ arises from the interference effect between the phonon $G$ band and the electronic Raman spectra (ERS) causing the Fano resonance~\cite{Hasdeo2014-fano,Hell2018}:
\begin{equation}
  \label{eq:intens}
  I(\omega_{\rm s} )= \left[A_G(\omega_{\rm s} )+A_{\rm ERS}(\omega_{\rm s} )\right]^2,
\end {equation}
where $A_G=\sum_\nu A_{\nu}$, is the $G$ phonon scattering amplitude
which consists of zone center ($\Gamma$ point) $\nu =$ LO and iTO
modes. $A_{\rm ERS}$ is the ERS scattering amplitude considering only the
first order process.  To obtain $A_G$ and $A_{\rm ERS}$, electronic energy
bands and wave functions for the electron-photon,
electron-phonon and electron-electron interactions have been obtained
from tight-binding (TB) method considering up to the three nearest
neighbors with TB parameters fitted from ARPES measurement (this
work) for each doping level. For simplicity, all Raman calculations are carried out in the $1\times 1$ BZ.

\subsection*{Acknowledgements}
NE, MH and AG acknowledge the ERC-grant no. 648589 'SUPER-2D' and funding from Quantum Matter and Materials as well as CRC1238 project A1. The stay at the Elettra synchrotron for ARPES experiments has been supported by Horizon 2020 EC programme under Grant Agreement No. 730872 (CALIPSOplus). H.H. acknowledges funding from DAAD for a stay in Germany. R.S. acknowledges JSPS KAKENHI (No. JP18H01810).

Figures 3a, 3b, 3c, and 3d were generated using the pybinding package for python3~\cite{pybinding}. The Mayavi software was used to generate Figure 3f~\cite{ramachandran2011mayavi}. 


\bibliography{temp}

\begin{thebibliography}{10}

\bibitem{Kopnin2011}
N.~B. Kopnin, T.~T. Heikkil\"a, G.~E. Volovik, High-temperature surface
  superconductivity in topological flat-band systems.
\newblock {\it Phys. Rev. B\/} {\bf 83}, 220503 (2011).

\bibitem{Leykam2018}
D.~Leykam, A.~Andreanov, S.~Flach, Artificial flat band systems: from lattice
  models to experiments.
\newblock {\it Advances in Physics: X\/} {\bf 3}, 1473052 (2018).

\bibitem{Sutherland1986}
B.~Sutherland, Localization of electronic wave functions due to local topology.
\newblock {\it Phys. Rev. B\/} {\bf 34}, 5208--5211 (1986).

\bibitem{Syozi1951}
I.~Syozi, Statistics of kagome lattice.
\newblock {\it Progress in Theoretical Physics\/} {\bf 6}, 306--308 (1951).

\bibitem{Mielke1991}
A.~Mielke, Ferromagnetism in the hubbard model on line graphs and further
  considerations.
\newblock {\it Journal of Physics A: Mathematical and General\/} {\bf 24},
  3311--3321 (1991).

\bibitem{Bilitewski2018}
T.~Bilitewski, R.~Moessner, Disordered flat bands on the kagome lattice.
\newblock {\it Phys. Rev. B\/} {\bf 98}, 235109 (2018).

\bibitem{Lieb1989}
E.~H. Lieb, Two theorems on the hubbard model.
\newblock {\it Phys. Rev. Lett.\/} {\bf 62}, 1201--1204 (1989).

\bibitem{Tasaki1992}
H.~Tasaki, Ferromagnetism in the hubbard models with degenerate single-electron
  ground states.
\newblock {\it Phys. Rev. Lett.\/} {\bf 69}, 1608--1611 (1992).

\bibitem{Drost2017}
R.~Drost, T.~Ojanen, A.~Harju, P.~Liljeroth, Topological states in engineered
  atomic lattices.
\newblock {\it Nature Physics\/} {\bf 13}, 668 (2017).

\bibitem{Slot2017}
M.~R. Slot, T.~S. Gardenier, P.~H. Jacobse, G.~C.~P. van Miert, S.~N. Kempkes,
  S.~J.~M. Zevenhuizen, C.~M. Smith, D.~Vanmaekelbergh, I.~Swart, Experimental
  realization and characterization of an electronic lieb lattice.
\newblock {\it Nature Physics\/} {\bf 13}, 672 (2017).

\bibitem{Zhong2016}
C.~Zhong, Y.~Xie, Y.~Chen, S.~Zhang, Coexistence of flat bands and dirac bands
  in a carbon-kagome-lattice family.
\newblock {\it Carbon\/} {\bf 99}, 65--70 (2016).

\bibitem{Marchenko2018}
D.~Marchenko, D.~V. Evtushinsky, E.~Golias, A.~Varykhalov, T.~Seyller,
  O.~Rader, Extremely flat band in bilayer graphene.
\newblock {\it Science Advances\/} {\bf 4}, eaau0059 (2018). DOI:
  10.1126/sciadv.aau0059.

\bibitem{Senkovskiy2018}
B.~V. Senkovskiy, D.~Y. Usachov, A.~V. Fedorov, T.~Marangoni, D.~Haberer,
  C.~Tresca, G.~Profeta, V.~Caciuc, S.~Tsukamoto, N.~Atodiresei, N.~Ehlen,
  C.~Chen, J.~Avila, M.~C. Asensio, A.~Y. Varykhalov, A.~Nefedov, C.~W{\"o}ll,
  T.~K. Kim, M.~Hoesch, F.~R. Fischer, A.~Gr{\"u}neis, Boron-doped graphene
  nanoribbons: Electronic structure and raman fingerprint.
\newblock {\it ACS Nano\/} {\bf 12}, 7571--7582 (2018).

\bibitem{faugeras16-rhombohedral}
Y.~Henni, H.~P. Ojeda~Collado, K.~Nogajewski, M.~R. Molas, G.~Usaj, C.~A.
  Balseiro, M.~Orlita, M.~Potemski, C.~Faugeras, Rhombohedral multilayer
  graphene: A magneto-raman scattering study.
\newblock {\it Nano Lett.\/} {\bf 16}, 3710--3716 (2016).

\bibitem{Pierucci2015}
D.~Pierucci, H.~Sediri, M.~Hajlaoui, J.-C. Girard, T.~Brumme, M.~Calandra,
  E.~Velez-Fort, G.~Patriarche, M.~G. Silly, G.~Ferro, V.~Souli{\'e}re,
  M.~Marangolo, F.~Sirotti, F.~Mauri, A.~Ouerghi, Evidence for flat bands near
  the fermi level in epitaxial rhombohedral multilayer graphene.
\newblock {\it ACS Nano\/} {\bf 9}, 5432--5439 (2015).

\bibitem{calandra17-rhombohedral}
B.~Pamuk, J.~Baima, F.~Mauri, M.~Calandra, Magnetic gap opening in
  rhombohedral-stacked multilayer graphene from first principles.
\newblock {\it Phys. Rev. B\/} {\bf 95}, 075422 (2017).

\bibitem{Henck2018}
H.~Henck, J.~Avila, Z.~Ben~Aziza, D.~Pierucci, J.~Baima, B.~Pamuk, J.~Chaste,
  D.~Utt, M.~Bartos, K.~Nogajewski, B.~A. Piot, M.~Orlita, M.~Potemski,
  M.~Calandra, M.~C. Asensio, F.~Mauri, C.~Faugeras, A.~Ouerghi, Flat
  electronic bands in long sequences of rhombohedral-stacked graphene.
\newblock {\it Phys. Rev. B\/} {\bf 97}, 245421 (2018).

\bibitem{Bistritzer2011}
R.~Bistritzer, A.~H. MacDonald, Moir{\'e} bands in twisted double-layer
  graphene.
\newblock {\it Proceedings of the National Academy of Sciences\/} {\bf 108},
  12233--12237 (2011).

\bibitem{Cao2018}
Y.~Cao, V.~Fatemi, S.~Fang, K.~Watanabe, T.~Taniguchi, E.~Kaxiras,
  P.~Jarillo-Herrero, Unconventional superconductivity in magic-angle graphene
  superlattices.
\newblock {\it Nature\/} {\bf 556}, 43 (2018).

\bibitem{rotenberg10-extended}
J.~L. McChesney, A.~Bostwick, T.~Ohta, T.~Seyller, K.~Horn, J.~Gonz\'alez,
  E.~Rotenberg, Extended van hove singularity and superconducting instability
  in doped graphene.
\newblock {\it Phys. Rev. Lett.\/} {\bf 104}, 136803 (2010).

\bibitem{Dresselhaus2002}
M.~S. Dresselhaus, G.~Dresselhaus, Intercalation compounds of graphite.
\newblock {\it Advances in Physics\/} {\bf 51}, 1-186 (2002).

\bibitem{takahashi16-cs}
J.~Kleeman, K.~Sugawara, T.~Sato, T.~Takahashi, {Enhancement of electron-phonon
  coupling in Cs-overlayered intercalated bilayer graphene}.
\newblock {\it Journal of Physics: Condensed Matter\/} {\bf 28}, 204001 (2016).

\bibitem{osgood2019}
Y.~Lin, G.~Chen, J.~T. Sadowski, Y.~Li, S.~A. Tenney, J.~I. Dadap, M.~S.
  Hybertsen, R.~M. Osgood, Observation of intercalation-driven zone folding in
  quasi-free-standing graphene energy bands.
\newblock {\it Phys. Rev. B\/} {\bf 99}, 035428 (2019).

\bibitem{Hell2018}
M.~Hell, N.~Ehlen, B.~V. Senkovskiy, H.~Hasdeo, A.~V. Fedorov, D.~Dombrowski,
  C.~Busse, T.~Michely, G.~Di~Santo, L.~Petaccia, R.~Saito, A.~Gr{\"u}neis,
  {Resonance Raman spectrum of doped epitaxial graphene at the Lifshitz
  transition}.
\newblock {\it Nano Letters\/} {\bf 18}, 6045 (\textbf{2018}).

\bibitem{Honerkamp2008}
C.~Honerkamp, Density waves and cooper pairing on the honeycomb lattice.
\newblock {\it Phys. Rev. Lett.\/} {\bf 100}, 146404 (2008).

\bibitem{Nandkishore2012-chiral}
R.~Nandkishore, L.~S. Levitov, A.~V. Chubukov, Chiral superconductivity from
  repulsive interactions in doped graphene.
\newblock {\it Nature Physics\/} {\bf 8}, 158-- (2012).

\bibitem{Profeta2012}
G.~Profeta, M.~Calandra, F.~Mauri, Phonon-mediated superconductivity in
  graphene by lithium deposition.
\newblock {\it Nat Phys\/} {\bf 8}, 131 (2012).

\bibitem{Chapman2016}
J.~Chapman, Y.~Su, C.~A. Howard, D.~Kundys, A.~N. Grigorenko, F.~Guinea, A.~K.
  Geim, I.~V. Grigorieva, R.~R. Nair, Superconductivity in ca-doped graphene
  laminates.
\newblock {\it Scientific Reports\/} {\bf 6}, 23254-- (2016).

\bibitem{ichinokura16-ca}
S.~Ichinokura, K.~Sugawara, A.~Takayama, T.~Takahashi, S.~Hasegawa,
  Superconducting calcium-intercalated bilayer graphene.
\newblock {\it ACS Nano\/} {\bf 10}, 2761-2765 (2016). PMID: 26815333.

\bibitem{Bianchi2010}
M.~Bianchi, E.~D.~L. Rienks, S.~Lizzit, A.~Baraldi, R.~Balog, L.~Hornek\ae{}r,
  P.~Hofmann, Electron-phonon coupling in potassium-doped graphene:
  Angle-resolved photoemission spectroscopy.
\newblock {\it Phys. Rev. B\/} {\bf 81}, 041403 (2010).

\bibitem{Haberer2013}
D.~Haberer, L.~Petaccia, A.~V. Fedorov, C.~S. Praveen, S.~Fabris, S.~Piccinin,
  O.~Vilkov, D.~V. Vyalikh, A.~Preobrajenski, N.~I. Verbitskiy, H.~Shiozawa,
  J.~Fink, M.~Knupfer, B.~B\"uchner, A.~Gr\"uneis, Anisotropic eliashberg
  function and electron-phonon coupling in doped graphene.
\newblock {\it Phys. Rev. B\/} {\bf 88}, 081401 (2013).

\bibitem{Fedorov2014}
A.~V. Fedorov, N.~I. Verbitskiy, D.~Haberer, C.~Struzzi, L.~Petaccia,
  D.~Usachov, O.~Y. Vilkov, D.~V. Vyalikh, J.~Fink, M.~Knupfer, B.~B\"uchner,
  A.~Gr\"uneis, Observation of a universal donor-dependent vibrational mode in
  graphene.
\newblock {\it Nat. Commun.\/} {\bf 5}, 4257 (2014).

\bibitem{nikolay16-barium}
N.~I. Verbitskiy, A.~V. Fedorov, C.~Tresca, G.~Profeta, L.~Petaccia, B.~V.
  Senkovskiy, D.~Y. Usachov, D.~V. Vyalikh, L.~V. Yashina, A.~A. Eliseev,
  T.~Pichler, A.~Gr{\"u}neis, Environmental control of electron-phonon coupling
  in barium doped graphene.
\newblock {\it 2D Materials\/} {\bf 3}, 045003 (2016).

\bibitem{dima18-epc}
D.~Y. Usachov, A.~V. Fedorov, O.~Y. Vilkov, I.~I. Ogorodnikov, M.~V. Kuznetsov,
  A.~Gr\"uneis, C.~Laubschat, D.~V. Vyalikh, Electron-phonon coupling in
  graphene placed between magnetic li and si layers on cobalt.
\newblock {\it Phys. Rev. B\/} {\bf 97}, 085132 (2018).

\bibitem{McMillan1968}
W.~L. McMillan, Transition temperature of strong-coupled superconductors.
\newblock {\it Phys. Rev.\/} {\bf 167}, 331--344 (1968).

\bibitem{Csanyi2005}
G.~Csanyi, P.~B. Littlewood, A.~H. Nevidomskyy, C.~J. Pickard, B.~D. Simons,
  The role of the interlayer state in the electronic structure of
  superconducting graphite intercalated compounds.
\newblock {\it Nature Physics\/} {\bf 1}, 42 (2005).

\bibitem{Smith2015}
R.~P. Smith, T.~E. Weller, C.~A. Howard, M.~P.~M. Dean, K.~C. Rahnejat, S.~S.
  Saxena, M.~Ellerby, Superconductivity in graphite intercalation compounds.
\newblock {\it Superconducting Materials: Conventional, Unconventional and
  Undetermined\/} {\bf 514}, 50--58 (2015).

\bibitem{Kanetani2012}
K.~Kanetani, K.~Sugawara, T.~Sato, R.~Shimizu, K.~Iwaya, T.~Hitosugi,
  T.~Takahashi, Ca intercalated bilayer graphene as a thinnest limit of
  superconducting c6ca.
\newblock {\it Proceedings of the National Academy of Sciences\/} {\bf 109},
  19610--19613 (2012).

\bibitem{Coraux2008}
J.~Coraux, A.~T. NDiaye, C.~Busse, T.~Michely, Structural coherency of graphene
  on ir(111).
\newblock {\it Nano Letters\/} {\bf 8}, 565-570 (2008).

\bibitem{Petrovic2013}
M.~Petrovic, I.~Srut~Rakic, S.~Runte, C.~Busse, J.~T. Sadowski, P.~Lazic,
  I.~Pletikosic, Z.-H. Pan, M.~Milun, P.~Pervan, N.~Atodiresei, R.~Brako,
  D.~Sokcevic, T.~Valla, T.~Michely, M.~Kralj, The mechanism of caesium
  intercalation of graphene.
\newblock {\it Nature Communications\/} {\bf 4}, 2772-- (2013).

\bibitem{Mak2011}
K.~F. Mak, J.~Shan, T.~F. Heinz, Seeing many-body effects in single- and
  few-layer graphene: Observation of two-dimensional saddle-point excitons.
\newblock {\it Phys. Rev. Lett.\/} {\bf 106}, 046401 (2011).

\bibitem{Mak2014}
K.~F. Mak, F.~H. da~Jornada, K.~He, J.~Deslippe, N.~Petrone, J.~Hone, J.~Shan,
  S.~G. Louie, T.~F. Heinz, Tuning many-body interactions in graphene: The
  effects of doping on excitons and carrier lifetimes.
\newblock {\it Phys. Rev. Lett.\/} {\bf 112}, 207401 (2014).

\bibitem{Ando1998}
T.~Ando, T.~Nakanishi, R.~Saito, Berry's phase and absence of back scattering
  in carbon nanotubes.
\newblock {\it J. Phys. Soc. Jpn.\/} {\bf 67}, 2857--2862 (1998).

\bibitem{rsaito-book}
R.~Saito, G.~Dresselhaus, M.~S. Dresselhaus, {\it Physical Properties of Carbon
  Nanotubes\/} (Imperial College Press, 1998).

\bibitem{CastroNeto2009}
A.~H. Castro~Neto, F.~Guinea, N.~M.~R. Peres, K.~S. Novoselov, A.~K. Geim, The
  electronic properties of graphene.
\newblock {\it Rev. Mod. Phys.\/} {\bf 81}, 109--162 (2009).

\bibitem{Saito2011-review}
R.~Saito, M.~Hofmann, G.~Dresselhaus, A.~Jorio, M.~S. Dresselhaus, Raman
  spectroscopy of graphene and carbon nanotubes.
\newblock {\it Advances in Physics\/} {\bf 60}, 413-550 (2011).

\bibitem{Ferrari2013-review}
A.~C. Ferrari, D.~M. Basko, Raman spectroscopy as a versatile tool for studying
  the properties of graphene.
\newblock {\it Nature Nanotechnology\/} {\bf 8}, 235-- (2013).

\bibitem{Tusche2016}
C.~Tusche, P.~Goslawski, D.~Kutnyakhov, M.~Ellguth, K.~Medjanik, H.~J. Elmers,
  S.~Chernov, R.~Wallauer, D.~Engel, A.~Jankowiak, G.~Sch{\"o}nhense,
  {Multi-MHz time-of-flight electronic bandstructure imaging of graphene on
  Ir(111)}.
\newblock {\it Appl. Phys. Lett.\/} {\bf 108}, 261602 (2016).

\bibitem{Bostwick2007}
A.~Bostwick, T.~Ohta, J.~L. McChesney, T.~Seyller, K.~Horn, E.~Rotenberg, Band
  structure and many body effects in graphene.
\newblock {\it The European Physical Journal Special Topics\/} {\bf 148}, 5--13
  (2007).

\bibitem{lazerri06-kohn}
M.~Lazzeri, F.~Mauri, Nonadiabatic kohn anomaly in a doped graphene monolayer.
\newblock {\it Phys. Rev. Lett.\/} {\bf 97}, 266407 (2006).

\bibitem{Ulstrup2016}
S.~Ulstrup, M.~Sch\"uler, M.~Bianchi, F.~Fromm, C.~Raidel, T.~Seyller,
  T.~Wehling, P.~Hofmann, Manifestation of nonlocal electron-electron
  interaction in graphene.
\newblock {\it Phys. Rev. B\/} {\bf 94}, 081403 (2016).

\bibitem{brus2011-dopedgraphene}
N.~Jung, B.~Kim, A.~C. Crowther, N.~Kim, C.~Nuckolls, L.~Brus, Optical
  reflectivity and raman scattering in few-layer-thick graphene highly doped by
  {K and Rb}.
\newblock {\it ACS Nano\/} {\bf 5}, 5708--5716 (2011).

\bibitem{parret13-rubidium}
R.~Parret, M.~Paillet, J.-R. Huntzinger, D.~Nakabayashi, T.~Michel, A.~Tiberj,
  J.-L. Sauvajol, A.~A. Zahab, In situ raman probing of graphene over a broad
  doping range upon rubidium vapor exposure.
\newblock {\it ACS Nano\/} {\bf 7}, 165-173 (2013). PMID: 23194077.

\bibitem{Kresse1996}
G.~Kresse, J.~Furthm\"uller, Efficient iterative schemes for ab initio
  total-energy calculations using a plane-wave basis set.
\newblock {\it Phys. Rev. B\/} {\bf 54}, 11169--11186 (1996).

\bibitem{Kresse1996a}
G.~Kresse, J.~Furthm{\"u}ller, Efficiency of ab-initio total energy
  calculations for metals and semiconductors using a plane-wave basis set.
\newblock {\it Computational Materials Science\/} {\bf 6}, 15--50 (1996).

\bibitem{Kresse1999}
G.~Kresse, D.~Joubert, From ultrasoft pseudopotentials to the projector
  augmented-wave method.
\newblock {\it Phys. Rev. B\/} {\bf 59}, 1758--1775 (1999).

\bibitem{Tresca2018}
C.~Tresca, N.~I. Verbitskiy, A.~Gr{\"u}neis, G.~Profeta, Ab initio study of the
  (2{$\times$}2) phase of barium on graphene.
\newblock {\it The European Physical Journal B\/} {\bf 91}, 165 (2018).

\bibitem{Hasdeo2014-fano}
E.~H. Hasdeo, A.~R.~T. Nugraha, M.~S. Dresselhaus, R.~Saito, Breit-wigner-fano
  line shapes in raman spectra of graphene.
\newblock {\it Phys. Rev. B\/} {\bf 90}, 245140 (2014).

\bibitem{Starnberg1997}
H.~I. Starnberg, H.~E. Brauer, V.~N. Strocov, Low temperature adsorption of cs
  on layered tis2 studied by photoelectron spectroscopy.
\newblock {\it Surface Science\/} {\bf 384}, L785--L790 (1997).

\bibitem{Murray2019}
C.~Murray, W.~Jolie, J.~A. Fischer, J.~Hall, C.~van Efferen, N.~Ehlen,
  A.~Gr\"uneis, C.~Busse, T.~Michely, Comprehensive tunneling spectroscopy of
  quasifreestanding ${\mathrm{mos}}_{2}$ on graphene on ir(111).
\newblock {\it Phys. Rev. B\/} {\bf 99}, 115434 (2019).

\bibitem{Ehlen2018a}
N.~Ehlen, J.~Hall, B.~V. Senkovskiy, M.~Hell, J.~Li, A.~Herman, D.~Smirnov,
  A.~Fedorov, V.~Yu~Voroshnin, G.~Di~Santo, L.~Petaccia, T.~Michely,
  A.~Gr{\"u}neis, {Narrow photoluminescence and Raman peaks of epitaxial MoS2
  on graphene/Ir(111)}.
\newblock {\it 2D Materials\/} {\bf 6}, 011006 (2018).

\bibitem{Petaccia2009}
L.~Petaccia, P.~Vilmercati, S.~Gorovikov, M.~Barnaba, A.~Bianco, D.~Cocco,
  C.~Masciovecchio, A.~Goldoni, Bad elph: A normal-incidence monochromator
  beamline at elettra.
\newblock {\it Nuclear Instruments and Methods in Physics Research Section A:
  Accelerators, Spectrometers, Detectors and Associated Equipment\/} {\bf 606},
  780--784 (2009).

\bibitem{Hell2018a}
M.~G. Hell, Y.~Falke, A.~Bliesener, N.~Ehlen, B.~V. Senkovskiy, T.~Szkopek,
  A.~Gr{\"u}neis, Combined ultra high vacuum raman and electronic transport
  characterization of large-area graphene on sio2.
\newblock {\it Phys. Status Solidi B\/} {\bf 255}, 1800456 (2018).

\bibitem{Faris1997-oxygen}
G.~W. Faris, R.~A. Copeland, Ratio of oxygen and nitrogen raman cross sections
  in the ultraviolet.
\newblock {\it Appl. Opt.\/} {\bf 36}, 2684--2685 (1997).

\bibitem{Grimme2006}
S.~Grimme, Semiempirical gga-type density functional constructed with a
  long-range dispersion correction.
\newblock {\it J. Comput. Chem.\/} {\bf 27}, 1787--1799 (2006).

\bibitem{Varykhalov2012}
A.~Varykhalov, D.~Marchenko, M.~R. Scholz, E.~D.~L. Rienks, T.~K. Kim,
  G.~Bihlmayer, J.~S\'anchez-Barriga, O.~Rader, Ir(111) surface state with
  giant rashba splitting persists under graphene in air.
\newblock {\it Phys. Rev. Lett.\/} {\bf 108}, 066804 (2012).

\bibitem{DalCorso2015}
A.~Dal~Corso, Clean ir(111) and pt(111) electronic surface states: A
  first-principle fully relativistic investigation.
\newblock {\it Surface Science\/} {\bf 637-638}, 106--115 (2015).

\bibitem{pybinding}
D.~Moldovan, M.~Andelkovic, F.~Peeters, {pybinding v0.9.4: a Python package for
  tight- binding calculations} (2017). {This work was supported by the Flemish
  Science Foundation (FWO-Vl) and the Methusalem Funding of the Flemish
  Government.}

\bibitem{ramachandran2011mayavi}
P.~Ramachandran, G.~Varoquaux, {Mayavi: 3D Visualization of Scientific Data}.
\newblock {\it Computing in Science \& Engineering\/} {\bf 13}, 40--51 (2011).

\end{thebibliography}
\bibliographystyle{ScienceAdvances}

\end{document}